# Potential energy barrier for proton transfer in compressed benzoic acid

Dominik Kurzydłowski[1*]

[1] Faculty of Mathematics and Natural Sciences, Cardinal Stefan Wyszyński University, 01-038 Warsaw, Poland;

Benzoic acid (BA) is a model system for studying proton transfer (PT) reactions. The properties of solid BA subject to high pressure (exceeding 1 kbar = 0.1 GPa) are of particular interest due to the possibility of compression-tuning of the PT barrier. Here we present a simulations aimed a evaluating the value of this barrier in solid BA in the 1 atm – 15 GPa pressure range. We find that pressure-induced shortening of the O···O contacts within the BA dimers leads to a decrease in PT barrier, and subsequent symmetrization of the hydrogen bond. However, this effect is obtained only after taking into account zero-point energy (ZPE) differences between BA tautomers and the transition state. The obtained results shed light on previous experiments on compressed benzoic acid, and indicate that a common scaling behavior with respect to O···O distance might be applicable for hydrogen-bond symmetrization in both organic and inorganic systems.

Proton transfer reactions lie at the heart of many chemical transformations, such as acid-base reactions, enzymatic processes or hydrogen transport in Earth's mantle.[1,2] Studying proton dynamics not only gives insight into these transformations, but also leads to a deeper understanding of the quantum effects that are strongly manifested during the PT process.[3,4] Benzoic acid, with its centrosymmetric dimers bound by strong hydrogen bonds, is considered a model system for studying proton transfer processes.[5,6] This molecule was studied with a multitude of experimental techniques, including x-ray and neutron scattering,[7–10] optical spectroscopy,[11–15] and nuclear magnetic resonance (NMR),[16–20] as well as *ab initio* modelling.[21–31]

Experiments and calculations were also conducted for solid benzoic acid at high pressure (exceeding 0.1 GPa) yielding information on the evolution of the hydrogen bond and its dynamics upon compression.[32–39] Because of the challenging nature of the high-pressure experiments ambiguity exists in the interpretation of the data obtained from these measurements. In particular Kang an co-workers reported the near symmetrization of hydrogen bonds in BA subject to pressure of 17 GPa.[38] However, their conclusions were supported by Density Functional Theory (DFT) calculations utilizing the relatively crude local-density approximation (LDA).

Here we present a computational study aiming at elucidating how high pressure influences the properties of solid benzoic acid, in particular the potential energy surface (PES) barrier towards double proton transfer within the benzoic acid dimers. We model this system in the pressure range from 1 atm to 15 GPa with the recently proposed SCAN functional,[40] which was shown to correctly reproduce the properties of liquid water, an archetypical hydrogen-bonded system.[41]

Periodic DFT calculations were conducted in VASP 6.2.1.[42,43] A planewave basis set with a 800 eV cut-off was used for the description of valence electrons, while the *1s* electrons of C and O were modelled with a projector-augmented-wave potential. Geometry optimization of lattice vectors and fractional atomic coordinates was performed with the use of the regularized version of the SCAN functional (r$^2$SCAN).[44] The convergence criterion for the electronic minimization was $10^{-8}$ eV per cell. Sampling of the Brillouin zone was done through a Monkhorst–Pack mesh with a $2\pi \times 0.05$ Å$^{-1}$ spacing. The r$^2$SCAN functional was also used for obtaining Γ-point vibration (phonon) frequencies with the finite-displacement method (0.005 Å displacement), as implemented in VASP. Zero-point energy was evaluated from Γ-point vibrational frequencies. Single-point energy calculations on the r$^2$SCAN-optimized structures were done with the use of the hybrid HSE06 functional.[45]

At ambient and high pressure benzoic acid crystallizes in the *P2$_1$/c* space group ($Z = 4$, one molecule in the asymmetric unit) with two equivalent dimers present in the unit cell.[10] The two tautomeric forms of these centrosymmetric dimers (A and B) are shown in Figure 1. They can convert to each other through a transfer of two protons. Due to crystal packing effects A and B tautomers are not energetically equivalent and at ambient conditions crystals containing the A form have an energy lower by 0.2 – 0.8 kJ/mol (throughout the text the energies are referenced to one dimer), as determined by various experimental techniques.[20] This is one of the smallest energy difference between tautomeric forms of carboxylic acid dimers – e.g. nearly 8 kJ per/mol is found for ibuprofen.[46]

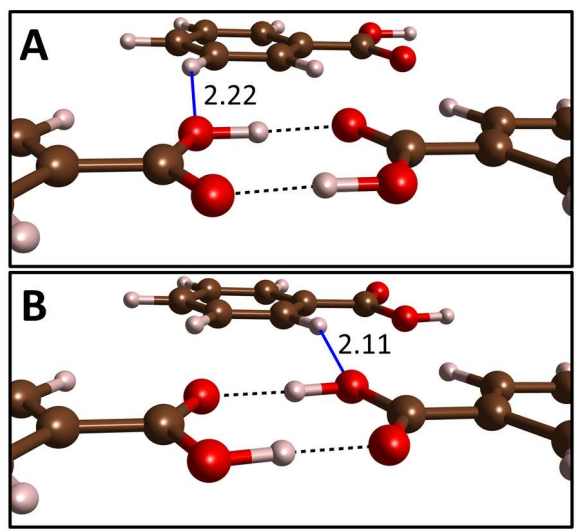

**Figure 1** Arrangement of molecules in the A (top) and B (bottom) tautomers of benzoic acid dimers. The length at 10 GPa of the interdimer O⋯H contacts formed by the hydroxyl oxygen atom (blue lines) are given in Å. Visualization performed with the VESTA software.[47]

It can be expected that the enthalpy difference between A and B tautomers, hereinafter referred to as asymmetry enthalpy, will increase upon compression of solid benzoic acid. To verify this hypothesis geometry optimization, utilizing the r$^2$SCAN functional, was performed in the 1 atm – 15 GPa pressure range for crystals containing solely A or B dimers. At ambient conditions the calculated asymmetry enthalpy (energy) is 0.6 kJ/mol (0.4 kJ/mol after including ZPE differences between the tautomers), in excellent agreement with experiment. As can be seen in Figure 2 (a) compression up to 10 GPa increases the enthalpy difference between the two tautomers to 2.6 kJ/mol (1.6 kJ/mol after including the ZPE correction). Above this pressure the ZPE-corrected asymmetry enthalpy starts to decrease due to the increase of the difference in zero-point energies between the two tautomers. The lower ZPE of the B form stems from a more elongated O-H bond (1.04 Å and 1.05 Å for A and B, respectively), and consequently lower O-H stretching frequencies (2309/2173 cm$^{-1}$ for A$_g$ modes for A/B). The lengthening of the O-H bond in B as compared to A can be linked with the destabilizing effect of the interdimer O···H contact of the hydroxyl atom which is shorter in this form (see Figure 1).

Our calculations not only correctly reproduce the enthalpy difference between A and B tautomers, but also yield geometry parameters in good agreement with experiment. As shown in Table S1 in Supporting Information, differences in lattice parameters and bond lengths do not exceed 2.2 % when comparing with both ambient pressure,[10] and high pressure data,[33,35] with the exception of the x-ray data of ref. [38] for which much larger discrepancies are found. We note however that the structural parameters reported in that study yield unphysically constant O···O distances at high pressure, as shown in Figure 2 (b). This points toward a likely error in the experimental structure determination.

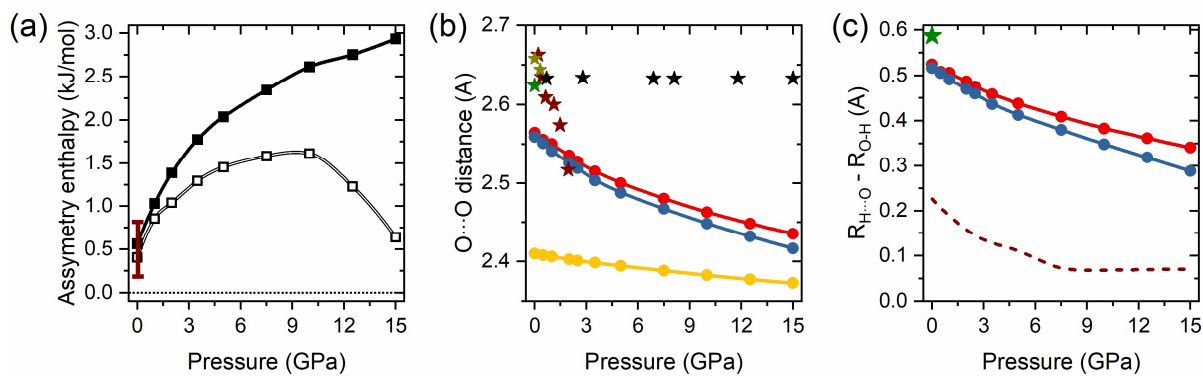

**Figure 2** (a) Value of the asymmetry enthalpy as a function of pressure (solid squares). Empty squares denote ZPE-corrected values, vertical bars denote the range of experimental values determined at 1 atm. (b) Calculated pressure evolution of the O···O distances within the hydrogen bond (dots) for A (red), B (blue), and the MEP transition state (yellow) structures of BA. Stars denote experimental values from ref. [10] (green), [33] (dark yellow), [35] (maroon), and [38] (black). (c) Calculated difference between the length of the O···H contact and the O-H bond (dots) for A (red) and B (blue) forms of benzoic acid dimers. Green star denotes the experimental value for A at 1 atm and 6K.[10] Dashed line denotes values obtained from previous LDA calculations.[38]

Upon compression the O⋯O contact is predicted to shorten in both forms of BA, as seen in Figure 2 (b), although this decrease is less abrupt that in the experimental data. Shorter O⋯O distance lead to strengthening of the hydrogen bond, and consequently the decrease in the difference between the O-H bond and H⋯O distance – see Figure 2 (c). Nevertheless, even at 15 GPa the system is far from hydrogen-bond symmetrization, in contrast to the results of previous LDA calculations.[38]

Pressure-induced strengthening of the hydrogen bond should lead to the decrease of the barrier for the double proton transfer within the benzoic acid dimers. In order to evaluate this effect a structure being a linear combination of A and B forms was constructed at each pressure point. The coefficients of this combination, both close to 0.5, where fine-tuned in such a way that the transferring protons in the resulting structure were position half-way between the O atoms. The difference in enthalpy between this structure and that of the A tautomer was used as a proxy for the PT barrier along the linear path (LP) trajectory. This trajectory is characterized by minimal displacement of heavy atoms, in particular the O⋯O distance remains nearly unchanged during PT.

The barrier along LP is larger than along the minimum energy path (MEP). However, it is postulated that the proton transfer trajectory follows a path of minimum action with a barrier height lying in between that of LP and MEP.[48] Therefore, the LP barrier can be treated as the upper bound, while the MEP barrier as a lower bound for the PT barrier. In order to gauge the value of the barrier along MEP, geometry optimization of the structures derived as a linear combination of A and B tautomers was performed. In this optimization, done at each pressure point with the use of the r$^2$SCAN functional, the fractional positions of the acidic hydrogen atoms were fixed while positions of other atoms, as well as the lattice parameters, were allowed to vary. Consequently, the MEP barrier was evaluated as the difference in enthalpy between the A form and the optimized structure. As can be seen in Figure 2 (b), the geometry optimization of the LP transition state leads to considerable shortening of the O⋯O contact at each pressure point. Even at ambient pressure the value of this distance (2.41 Å) is considerably smaller that O⋯O distances observed so far for carboxylic acid dimers.[49]

Previous studies indicated that the correct description of PT barriers by DFT methods requires the use of hybrid functionals which incorporate a portion of Hartree-Fock exchange.[50,51] Therefore the abovementioned enthalpy differences for LP and MEP trajectories were evaluated from single-point energy calculations performed on r$^2$SCAN-optimized structures with the use of the hybrid HSE06 functional.[45]

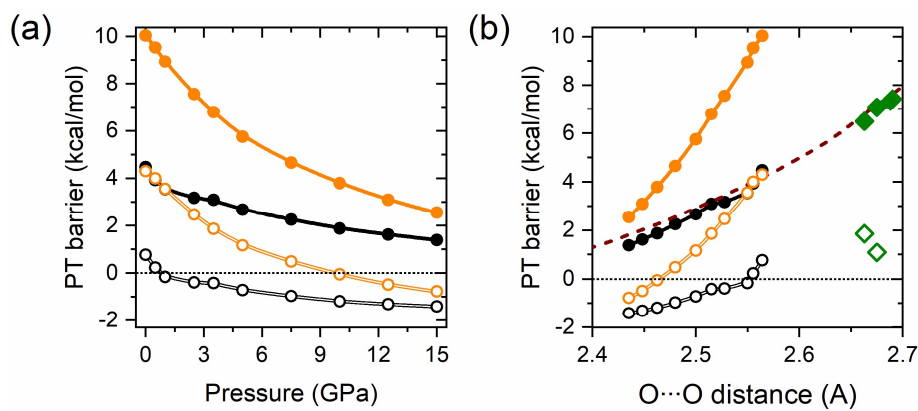

**Figure 3** (a) Pressure dependence of the proton transfer barrier (dots) along the MEP (black) and LP (orange) trajectories. Open symbols denote ZPE-corrected values. (b) The same quantities as a function of the O···O distance in the A tautomer. Diamonds denote theoretical values of the MEP barrier for isolated BA dimers obtained with the B3LYP functional,[25,27,28] and the MP2 method.[24] Open symbols denote ZPE-corrected values. Dashed maroon line indicates the dependence of the MEP barrier on the O···O distance in a tyrosine triad model obtained with the B3LYP-D3 functional.[4]

The PT barriers obtained for solid BA with the methodology described above are presented in Figure 3 as a function of pressure and the O···O distance in the A tautomer. The MEP barrier at 1 atm (4.5 kcal/mol) is smaller than theoretical values previously reported for isolated dimers (6.5 – 7.4 kcal/mol).[4,24,25,27,28] Inspection of Figure 3 (b) reveals that this can be traced back to the shorter O···O distance in the r$^2$SCAN-optimized structure. The MEP barrier decreases upon compression due to the shortening of the O···O distance. Interestingly, the dependence of the barrier on that distance is very similar to that obtained with the hybrid B3LYP functional for a model of hydrogen-bonded tyrosine residues within the active site of the enzyme ketosteroid isomerase.[4] This indicates that the energetics of the PT process may be similar regardless of the environment surrounding the hydrogen bond, as noted in a recent study.[52]

At ambient pressure the LP barrier (10.1 kcal/mol) is substantially larger than MEP barrier, but due to a more pronounced dependence on the O···O distance the LP barrier it decreases much faster with increasing pressure, as shown in Figure 3 (a). Even at 15 GPa the barriers for both the LP (2.6 kcal/mol) and MEP (1.5 kcal/mol) trajectories are still substantially larger than thermal energy at room temperature (0.6 kcal/mol). However, inclusion of ZPE corrections leads to a substantial lowering of the barriers at all pressures, as evident from Figure 3. This is due to the fact that the symmetric (A$_g$) O-H stretching modes have imaginary frequencies in the transition states. The ZPE-corrected value of the MEP barrier is only 0.7 kcal/mol at 1 atm, and is predicted to vanish upon compression to 0.8 GPa. Above this pressure, which corresponds to an O···O distance in the A tautomer of 2.55 Å, benzoic acid crystals with a symmetric hydrogen bond have a lower ZPE-corrected enthalpy than either the A or B tautomers. For the LP trajectory the ZPE-corrected barrier (4.5 kcal/mol at 1 atm) becomes zero at 9.7 GPa which corresponds to an O···O distance in A of 2.46 Å. We note that range of O···O distances spanned by the values at which the MEP and LP barriers vanish are close to the 2.5 Å value recently proposed as the borderline for ZPE overcoming the PT barrier height in hydrogen-bonded systems.[53] Moreover the distance at which the LP barrier becomes null (2.46 Å) is close to the critical O···O distance (2.44 Å) at which hydrogen-bond symmetrization is proposed for a number of inorganic systems.[52]

The results indicate that pressure-induced symmetrization of the hydrogen bond should occur in solid BA, but retrieving this effect requires taking into account the quantum nature of the system. Calculations suggest that this process should occur between 0.8 and 9.7 GPa. Taking into account the 2.2 % underestimation of the length of O···O contacts in both tautomers, leading to the underestimation of PT barriers, the experimental pressure at which the MEP and LP barriers vanish can be estimated as closer to 6 and 20 GPa, respectively. Nonetheless, even at lower pressure the barrier for the PT transition, which as mentioned earlier should lie in between the MEP and LP barriers, should become comparable to thermal energy. As a result, structural data obtained from experiments yielding time-averaged information (*e.g.* x-ray diffraction) should contain a considerable "admixture" from the PT transition state, particularly in the case of room-temperature experiments. This, together with the fact that the MEP transition state is predicted to exhibit very short O···O contacts in, might explain the dramatic decrease of O···O distances in the room-temperature measurements.[35]

Simulation presented in this work show how pressure can be used to tune the barrier of the proton transfer reaction in solid benzoic acid. It is shown that compression induces a barrierless transfer of proton, and that the vanishing of the barrier is strongly connected with differences in the ZPE energy. This indicates that quantum effects retain their important role in PT processes even at large compression. The obtained results shed light on previous experiments on compressed benzoic acid, in particular the drastic reduction of the O···O distance with pressure. The current findings are also of importance for other high-pressure studies on strongly hydrogen bonded systems,[52] most notably water.[54–56] In particular, it seems that the recently proposed critical O···O distance for hydrogen bond symmetrization (ref. [52]) might be common also to organic compounds. Benzoic acid is of special interest in this context, as this distance can obtained for this compound at relatively low pressure (p < 20 GPa). Therefore new experimental investigation aimed at establishing the proton dynamics in this compound are required.

**Acknowledgments**: D.K. acknowledges the support from the Polish National Science Centre (NCN) within the OPUS (grant number UMO-2016/23/B/ST4/03250). This research was carried out with the support of the Interdisciplinary Centre for Mathematical and Computational Modelling (ICM), University of Warsaw, within allocation no. GA83-26.

**Notes**: The author declares no competing financial interest.

**Supporting Information Available:** Comparison of the calculated and experimental geometry of BA crystals at ambient and high pressure

# Supporting Information

**Table S 1** Comparison of the calculated and experimental geometry of BA crystals. Bond distances and cell vector lengths are given in Å, volume in Å$^3$. Percentage difference between experimental data and r$^2$SCAN calculations is given in parenthesis. Theoretical values are reported for the A tautomeric form.

| p = 0 GPa | Exp (ref. [10]) | Theory | p = 0.3 GPa | Exp (ref. [33]) | Theory |
|---|---|---|---|---|---|
| a | 5.401 | 5.483 (+1.5) | a | 5.315 | 5.405 (+1.7) |
| b | 5.004 | 5.037 (+0.7) | b | 4.987 | 4.979 (–0.2) |
| c | 21.879 | 21.945 (+0.3) | c | 21.630 | 21.631 (0.0) |
| V | 584.8 | 602.92 (+3.1) | V | 566.3 | 575.4 (+1.6) |
| $R_{O \cdots O}$ | 2.623 | 2.564 (–2.2) | | | |
| $R_{O-H}$ | 1.019 | 1.021 (+0.2) | | | |
| $R_{C-O}$ | 1.320 | 1.316 (–0.3) | | | |
| $R_{C=O}$ | 1.233 | 1.242 (+0.7) | | | |
| **p = 2.2 GPa** | **Exp (ref. [35])** | **Theory** | **p = 15 GPa** | **Exp (ref. [38])** | **Theory** |
| a | 5.077 | 5.094 (+0.3) | a | 4.768 | 4.608 (–3.4) |
| b | 4.893 | 4.852 (–0.8) | b | 4.692 | 4.597 (–2.0) |
| c | 21.1 | 20.998 (–0.5) | c | 19.983 | 19.431 (–2.8) |
| V | 515.0 | 509.4 (–1.1) | V | 428.2 | 396.6 (–7.4) |